\documentclass[superscriptaddress,twocolumn,showkeys]{revtex4}
\usepackage{textcomp}
\usepackage{amsmath}
\usepackage{amssymb}
\usepackage{graphicx}
\usepackage{epstopdf}
\usepackage{float}

\usepackage[a4paper,bindingoffset=0.2in,left=0.7in,right=0.7in,top=1in,bottom=1in,footskip=.25in]{geometry}
\usepackage{color}
\begin{document}
\title{ Quark stars admixed with dark matter}          
\author{Payel Mukhopadhyay}
\email{pm12ms010@iiserkol.ac.in}
\affiliation{Indian Institute of Science Education and Research, Kolkata, Mohanpur, 741252, Nadia, West Bengal, India}
\affiliation{Institut f\"ur Theoretische Physik, J.W.Goethe Universit\"at, Max Von Laue-Stra\ss e 1, D-60438 Frankfurt Am Main, Germany}
\author{J\"urgen Schaffner-Bielich}
\email{schaffner@astro.uni-frankfurt.de}
\affiliation{Institut f\"ur Theoretische Physik, J.W.Goethe Universit\"at, Max Von Laue-Stra\ss e 1, D-60438 Frankfurt Am Main, Germany}

\date{November 1, 2015 }

\begin{abstract}
Compact stars consisting of massless quark matter and fermionic dark matter are studied by solving the Tolman-Oppenheimer-Volkoff equations for two fluids separately. Dark matter is further investigated by incorporating inter-fermionic interactions among the dark matter particles. The properties of stars made of quark matter particles and self-interacting and free dark matter particles are explored by obtaining their mass-radius relations. The regions of stability for such a compact star are determined and it is demonstrated that the maximum stable total mass of such a star decreases approximately linearly with increasing dark matter fraction.

\end{abstract}
\keywords{compact stars, quark matter, dark matter, mass-radius relations, equation of state}
\maketitle

\section{Introduction}
A quark star is a hypothetical compact star and consists of self-bound strange quark matter (SQM) \cite{Witten:1984abc,Itoh:1970abc,Farhi:1984abc,Weber:2005abc,Ivanenko:1965abc}. The existence of quark stars is controversial and its equation of state is also uncertain. 

One of the popular models for the equation of state of the quark star is the so called MIT Bag model \cite{Chodos:1974abc}. The model is often used for describing cold and massless (strange) quark matter \cite{Haensel:1986abc,Alcock:1986abc}. Standard values for the MIT bag constant are around \textit{$B^{1/4}$} = 145 MeV as follows from fits to hadron masses \cite{Haensel:1986abc}, which results in maximum masses of about 2.0$ M_{\odot}$ at a radius of about 11 km \cite{Witten:1984abc,Haensel:1986abc} , which are actually very close to the ones of realistic neutron star models. 

Dark matter stars are modelled in our work as a free or self-interacting fermion gas at zero temperature. The possible candidates for dark matter particles are a type of fermion predicted in extensions of the standard model including supersymmetric particles, the neutralino, the gravitino and the axino \cite{Baltz:2004abc}. In our discussion, we consider dark matter to be made of fermionic particles with a mass of 100 GeV, the classical WIMP mass scale. We assume that the dark matter particles cannot self-annihilate as in asymmetric dark matter (ADM) \cite{Kaplan:1992abc,Nussinov:1985abc}.

Self-annihilating WIMP dark matter with masses above a few GeV accreted onto neutron stars may trigger a conversion of most of the star into a strange star \cite{Perez:2010abc} or the accreted dark matter may significantly affect the kinematical properties of the compact star \cite{Perez:2012abc}. Constraints on the properties of dark matter candidates can be obtained from stars which can accrete asymmetric dark matter in its lifetime and then collapse into a neutron star \cite{Kouvaris:2011abc}. Constraints on the mass of dark matter candidates can also be obtained by the possible collapse of compact stars due to dark matter accretion \cite{Goldman:2013qla,Kouvaris:2015rea}. The cooling process of compact objects can be affected by the capture of dark matter which can annihilate the star \cite{Lavallaz:2010abc}. Recent studies have been done to explore compact stars with non self-annihilating dark matter to analyze the gravitational effects of dark matter on the stellar matter under intense conditions \cite{Li:2012abc,Leung:2011abc,Sandin:2009abc,Xiang:2014abc}. In these studies, masses of dark matter in the GeV range has been assumed. Studies have also been performed to investigate the compact objects formed due the admixture of neutron star matter and dark matter \cite{Laura:2015abc} leading to the possibilities for new stable solutions of compact stars with planet like masses. To study of the effect of dark matter on compact objects is therefore of great interest. If quark stars do exist in nature, they can also accumulate dark matter and hence their properties might change. This accumulation will lead to various changes in the mass-radius relation of a quark star which is studied in this work. 

This paper is organised as follows: in section 2, we shortly discuss about the two fluid TOV equations. In section 3, we discuss the equations of state for both quark matter and fermionic dark matter and discuss the general scaling solutions for these stars. In section 4, we present the numerically obtained results (mass-radius relations) for quark matter stars by solving the TOV equations. In section 5, TOV equations are solved for dark matter composed of both strongly self-interacting and free fermions and their corresponding mass-radius relations are obtained. Section 6 is dedicated to the numerical solutions of two fluid TOV equations namely, dark matter and quark matter which are coupled together only by gravity. We demonstrate that the maximum mass of the quark star admixed with dark matter reduces due to the presence of dark matter and decreases in a linear fashion in case in strongly self-interacting dark matter fermions  while for free dark matter, the maximum mass remains almost unaffected. Finally, in section 7, we summarize our findings and discuss our results.

Throughout the paper, we use natural units where \textit{c}=$\hbar$=1, \textit{c} being the speed of light and $\hbar$ is the reduced Planck's constant. 
   
\section{Two fluid Tolmann-Oppenheimer-Volkoff equations}

Since our aim is to see the properties of a quark star admixed with dark matter, we need the TOV equations for two fluids admixed with each other. There will be a hydrostatic equilibrium condition for each of the two fluids and the fact that there is only gravitational interaction between them will be encoded in the metric describing the system. The two fluid TOV equations that we use here are \cite{Sandin:2009abc,Laura:2015abc}:

\begin{equation}
\begin{split}
\emph{\(\frac{dp_{1}}{dr}\)} &=-\emph{\(\frac{GM(r)\rho_{1}(r)}{r^{2}}\)}\left(1+\emph{\(\frac{p_{1}(r)}{\rho_{1}(r)}\)}\right)\times \\
& \left(1+4\pi r^{3} \emph{\(\frac{(p_{1}(r)+p_{2}(r))}{M(r)}\)}\right)\left(1-2G\emph{\(\frac{M(r)}{r}\)}\right)^{-1}
\end{split}
\end{equation}

\begin{equation}
\begin{split}
\emph{\(\frac{dp_{2}}{dr}\)} &=-\emph{\(\frac{GM(r)\rho_{2}(r)}{r^{2}}\)}\left(1+\emph{\(\frac{p_{2}(r)}{\rho_{2}(r) }\)}\right)\times\\
&\left(1+4\pi r^{3} \emph{\(\frac{(p_{1}(r)+p_{2}(r))}{M(r)}\)}\right)\left(1-2G\emph{\(\frac{M(r)}{r}\)}\right)^{-1}
\end{split}
\end{equation}
\begin{equation}
\emph{\(\frac{dM_{1}}{dr}\)}=4\pi r^{2} \rho_{1}(r)
\end{equation} 
\begin{equation}
\emph{\(\frac{dM_{2}}{dr}\)}=4\pi r^{2} \rho_{2}(r)
\end{equation} 
\begin{equation}
M(r)=M_{1}(r)+M_{2}(r)
\end{equation}  

Here \textit{M(r)} represents the total mass at radius \textit{r}, \textit{$p_{1}$}, \textit{$p_{2}$}, $\rho_{1}$ and $\rho_{2}$ are the pressures and densities of the fluids 1 and 2 respectively. We could separate out the hydrostatic equilibrium condition for the two stars into equations (1) and (2) because the interaction acts only through gravity and nothing else. The gravitational interaction is taken into account because of the fact that the  mass that is considered in the equation \textit{M(R)} is the total mass of both the fluids at radius \textit{r} which means each of the fluid attracts the other gravitationally. The equations for the conservation of mass for the two fluids remains the same as that for individual fluids.

For solving the two fluid TOV equations, we need proper boundary conditions. \textit{$M_{1}(0)$} and {$M_{2}(0)$} must be equal to zero at \textit{r}=0. Central pressures for the two fluids are calculated from the central densities given as the initial condition using the respective equation of states for the two fluids. Then the two TOV equations are solved together simultaneously and we obtain either \textit{$R_{1}$} or \textit{$R_{2}$} as the radius of the complete star depending on which fluid ends up having a larger radius. The radius of the individual fluids occur at those points where the individual pressures drop down to zero.

\section{Equations of state for quark matter and free and self-interacting dark matter}
The equations for state for quark matter is discussed using MIT Bag model. The EOS for free dark matter particles along with strongly self-interacting dark matter particles is briefly described using statistical mechanics of free and self-interacting fermions. Scaling relations for quark stars and dark matter stars is also discussed.

\subsection{ Equation of state for quark matter}

The MIT Bag equation of state \cite{Haensel:1986abc,Alcock:1986abc} is taken as the equation of state for quark matter in our work. In this model, the quarks are assumed to be made of free fermions constrained within a bag with a vaccumm pressure that keeps the particles within the bag. The MIT Bag equation of state is:
\begin{equation}
p = \emph{\(\frac{1}{3}\)}(\epsilon -4B)
\end{equation}
Here, \textit{p} denotes the pressure, $\epsilon$ denotes the energy density and \textit{B} is the Bag constant whose standard accepted values are around \textit{B}\textsuperscript{1/4} = 145 MeV or \textit{B}\textsuperscript{1/4}= 200 MeV \cite{Haensel:1986abc}. Note that the equation of state for a cold gas of interacting massless quarks within perturbative quantum chromodynamics can be approximated by a similar form of equation of state as the MIT Bag model \cite{Fraga:2001abc}.

\subsection{Equation of state for free and self-interacting dark matter fermions  }
Dark matter will be assumed to be made of fermions of mass 100 GeV. For a study of compact fermionic stars, we refer to some recent papers \cite{Sagert:2006abc,Macher:2005abc,Silbar:2004abc}. The equation of state for a gas of free fermions can be calculated via explicit expressions for energy density ($\epsilon$) and pressure (\textit{p}) \cite{Haensel:2007abc}
\begin{equation}\label{xx}
\small
\begin{split}
\epsilon &=\emph{\(\frac{1}{\pi^{2}}\)} \int_{0}^{k_{F}} k^2 \sqrt{m_f^2 + k^2} dk\\
&=\emph{\(\frac{m_f^4}{8\pi^{2}}\)}[(2z^3+z)\sqrt{1+z^2}-sinh^{-1}(z)] 
\end{split}
\end{equation}
\begin{equation}\label{xx}
\small
\begin{split}
p &=\emph{\(\frac{1}{3\pi^{2}}\)} \int_{0}^{k_{F}} \emph{\(\frac{k^4}{\sqrt{m_f^2 + k^2}}\)} dk\\
&=\emph{\(\frac{m_f^4}{24\pi^{2}}\)}[(2z^3-3z)\sqrt{1+z^2}+3sinh^{-1}(z)] 
\end{split}
\end{equation}
where \textit{z}=$k_{F}$/$m_{f}$ is the dimensionless Fermi momentum.

Similarly, the interactions between the fermions is modelled by considering the simplest two-body interactions between fermions. The repulsion amongst the fermions constituting the dark matter star has been modelled by considering the interaction energy density to be proportional to \textit{$n^2$} \cite{Gaurav:2006abc,Pratik:2009abc,Rainer:2010abc} to the lowest order approximation, where \textit{$n$} is the number density of fermions. The resulting equation of state has been calculated in reference \cite{Gaurav:2006abc}:

\begin{equation}\label{xx}
\small
\begin{split}
\emph{\(\frac{\epsilon}{m_f^{4}}\)} &= \emph{\(\frac{1}{8\pi^{2}}\)}[(2z^3+z)\sqrt{1+z^2}-sinh^{-1}(z)] \\& + [\left(\emph{\(\frac{1}{3\pi^{2}}\)}\right)^2 y^2z^6]
\end{split}
\end{equation}
\begin{equation}\label{xx}
\small
\begin{split}
\emph{\(\frac{p}{m_f^{4}}\)} &=\emph{\(\frac{1}{24\pi^{2}}\)}[(2z^3-3z)\sqrt{1+z^2}+3sinh^{-1}(z)]\\ &+[\left(\emph{\(\frac{1}{3\pi^{2}}\)}\right)^2 y^2z^6]
\end{split}
\end{equation}
where \textit{z}=$k_{F}$/$m_{f}$ is again the dimensionless Fermi momentum and \textit{y} is the dimensionless interaction strength. Also, \textit{y}=$m_{f}$/$m_{I}$ where \textit{$m_I$} is the scale of interaction.
The mass of the fermions $m_{f}$ used for self-interacting and free dark matter has been taken as 100 GeV and it is assumed that they don't self annihilate \cite{Kaplan:1992abc,Nussinov:1985abc}. The value of the interaction strength \textit{y} determines whether the self-interactions are weak or strong. For example, neutralinos, that form a candidate for WIMP dark matter and are in the mass range of 100 GeV \cite{Bottino:2005abc} can have weak self-interactions with \textit{y} $\sim$ 0.1 or they may be strongly self-interacting with \textit{y} $\sim$ $10^3$ where the strong interaction scale corresponds to $\Lambda_{QCD}$ $\simeq$ 200 MeV. In our discussions, we would focus on two situations, one for free fermionic dark matter matter with \textit{y}=0 and the other for strongly self-interacting dark matter with \textit{y}=$10^3$. For a reference of self-interacting fermionic stars and the corresponding interaction strengths we refer to \cite{Gaurav:2006abc}.

\subsection{Scaling relations for quark matter and dark matter }

We generally scale dimensional quantities to dimensionless ones in order to represent any arbitrary mass configuration of a star in a single graph. From equation (6), it is clear that if we scale the energy density and pressure values by four times the Bag Constant (4\textit{B}), the EOS reduces to a dimensionless  form \cite{Witten:1984abc,Haensel:1986abc} the corresponding total mass and radius of the star would then be scaled by $\sqrt{4B}$. Similarly , for the dark matter fermionic particles, it is a natural choice to scale the pressure and energy density by the fermion mass $m_{f}^4$ which will again make the equations dimensionless. The scaling relations for quark matter are $\epsilon^{\prime}_{quark}$ = $\epsilon_{quark}$/(4\textit{B}), \textit{$p^{\prime}_{quark}$} = \textit{$p_{quark}$}/(4\textit{B}), \textit{$M^{\prime}_{quark}$} = \textit{$M_{quark}$}/(2\textit{$\sqrt B$}), \textit{$R^{\prime}_{quark}$} = \textit{$R_{quark}$}/(2\textit{$\sqrt B$}). The corresponding relations for fermions are $\epsilon^{\prime}_{f}$ = $\epsilon_{f}$/\textit{$m_{f}^4$}, \textit{$p^{\prime}_{f}$} = \textit{$p_{f}$}/\textit{$m_{f}^4$}, \textit{$M^{\prime}_{f}$} = \textit{$M_{f}$}/\textit{a}, \textit{$R^{\prime}_{f}$} = \textit{$R_{f}$}/\textit{b} where \textit{a} = \textit{$M_{p}^{3}$}/\textit{$m_{f}^{2}$} and \textit{b} = \textit{$M_{p}$}/\textit{$m_{f}^{2}$} where \textit{$M_{p}$} is the Planck Mass ( \textit{G} = \textit{$M_{p}^{-2}$} ). For detailed derivations of the scaling relations we refer to \cite{Pratik:2009abc}.

\section{ Solving TOV equations for quark matter star}

Numerical solutions to the mass-radius relations of quark stars can be found in literature \cite{Andreas:2015abc}. In our nomenclature \textit{$M_{quark}$} and \textit{$R_{quark}$} represents the mass and radius of the quark star respectively. The curve is shown in Fig. 1 for two different Bag values.

\begin{figure}[htb!]
\centering{}
\includegraphics[height=7.5cm,width=3.5in]{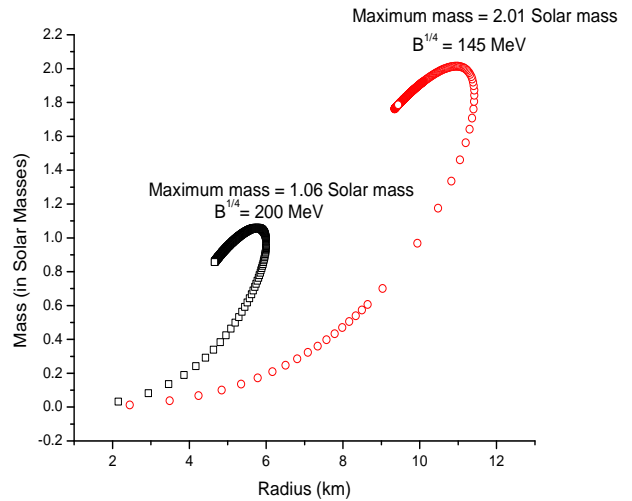}
\caption{Mass (\textit{$M_{quark}$}) vs. Radius (\textit{$R_{quark}$}) curve for quark stars for two different Bag values.}
\end{figure} 

Upto a certain point mass increases with the radius reaching a maximum value of mass at a certain value of radius after which the mass starts decreasing, the star starts becoming unstable from this point.The maximum stable mass for $\textit{B}^{1/4}$ = 145 MeV is about 2.01 $ M_{\odot}$ and the corresponding radius of around 11 km. While $\textit{B}^{1/4}$=200 MeV gives a maximum mass of about 1.06 $M_{\odot}$ with the radius being around 5.8 km. Quark stars are incompressible stars and form a self bound system \cite{Itoh:1970abc,Ivanenko:1965abc}. 

\section{Solving TOV equation numerically for free and Strongly self-interacting dark matter particles}

\subsection{Solutions for free fermionic dark matter  }
We first consider dark matter made of free fermionic particles with 100 GeV mass. Single fluid TOV equations are solved taking (7) and (8) as the equation of state. \textit{$M_{dark}$} and \textit{$R_{dark}$} represents the mass and radius of the dark matter star composed of free fermions respectively. The resulting mass-radius curve is plotted in Fig. 2.
\begin{figure}[htb!]
\centering{}
\includegraphics[height=7.5cm,width=3.5in]{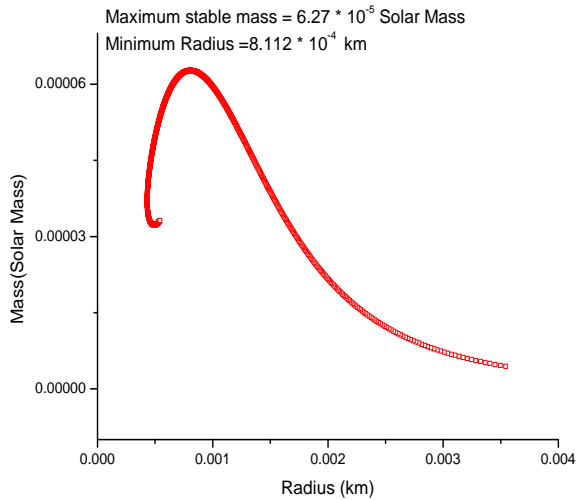}
\caption{Plot of the mass (\textit{$M_{dark}$}) vs. radius (\textit{$R_{dark}$}) for a free gas of fermions of mass 100 GeV at zero temperature. This graph corresponds to the mass-radius curve of the dark matter composed of free fermionic particles.  }
\end{figure} 

From the graph, we see that the mass at first increases with a decrease in radius for increasing central energy density values, reaches a maximum and then starts decreasing. Stellar configurations to the right side of the maximum masse are stable whereas those on the left side are unstable. The maximum stable mass for the dark matter star made of free fermions comes out to be 6.27$\times$ $10^{-5}$ $M_{\odot}$ with a radius of 0.81 meters.

\subsection{Solutions for strongly self-interacting dark matter}
The TOV equations are solved for strongly self-interacting dark matter particles (\textit{y} = $10^3$) of mass 100 GeV. The equation of state used are (9) and (10).  \textit{$M_{int}$} and \textit{$R_{int}$} represents the mass and radius of the dark matter star composed of strongly self-interacting fermions respectively.  The resulting mass-radius curve is plotted in Fig 3.
\begin{figure}[htb!]
\centering{}
\includegraphics[height=7.5cm,width=3.5in]{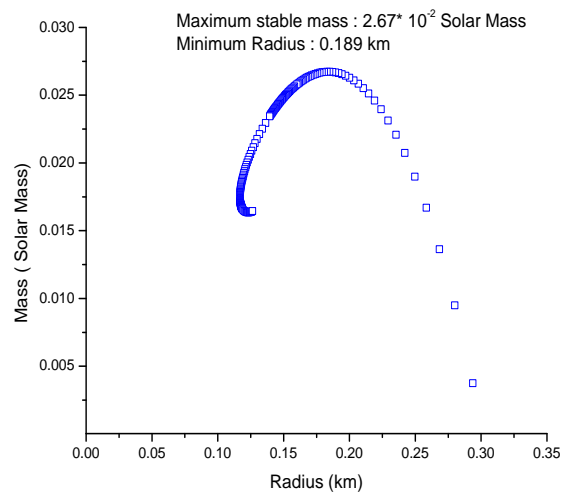}
\caption{Plot of  mass \textit{$M_{int}$} vs. radius \textit{$R_{int}$}  for strongly self-interacting dark matter fermionic particles (\textit{y} = $10^{3}$). }
\end{figure}

For strong self-interaction, the mass and radius are much larger compared to free fermions and hence the maximum mass and the minimum radius is about 1000 times larger. From the curve it is observed that for very low central densities of dark matter particles, i.e the tail of the graph, the rate of increase of mass with decreasing radius is much higher as compared to the free dark matter particle case discussed in the previous subsection. The maximum mass and the minimum radius for the self-interacting dark matter star   turns out to be 2.67 $\times$ $10^{-2}$ $M_{\odot}$ and 0.189 km respectively, larger than for the non interacting case due to repulsive forces between the dark matter particles.

\section{ Solution of TOV equation for an admixture of quark matter and dark matter }
Nomenclature used is \textit{$M_{quark}$} and \textit{$R_{quark}$} for the mass and radius of quark matter, \textit{$M_{dark}$} and \textit{$R_{dark}$} for the mass and radius of the star composed of free dark matter particles, \textit{$M_{int}$} and \textit{$R_{int}$} for strongly self-interacting dark matter star. $\epsilon_{0,quark}$, $\epsilon_{0,dark}$ and $\epsilon_{0,intdark}$ represents the central energy densities of quark matter, dark matter made of free fermions and dark matter made of strongly self-interacting fermions respectively.

\subsection{Solution for combination of quark matter and free dark matter particles}

\begin{figure}[htb!]
\centering{}
\includegraphics[height=7.5cm,width=3.5in]{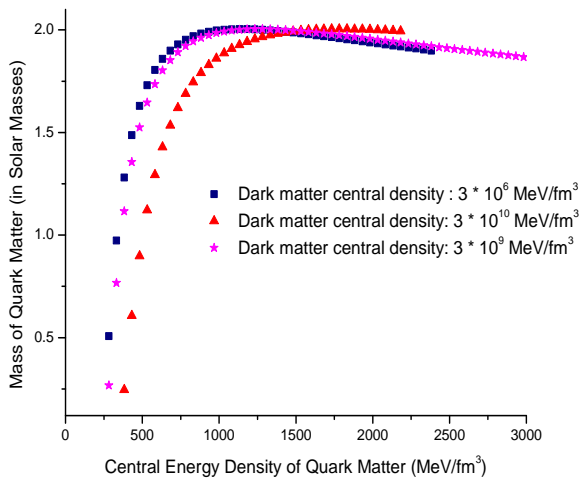}
\caption{Plot of mass (\textit{$M_{quark}$}) vs. central energy density ($\epsilon_{0,quark}$) of quark matter for three different central energy densities of dark matter ($\epsilon_{0,dark}$).}
\end{figure} 

The two fluid TOV equations (1), (2), (3) and (4) are solved for a mixture of quark matter with MIT Bag model by taking the bag value to be \textit{$B^{1/4}$}= 145 MeV and dark matter composed of free fermionic particles of mass 100 GeV. We start with the initial given central energy densities for the two components and compute the corresponding central pressures using the EOS for the respective fluids (eqn. (6) for quark matter and (7) and (8) for dark matter).
\begin{figure}[htb!]
\centering{}
\includegraphics[height=7.5cm,width=3.5in]{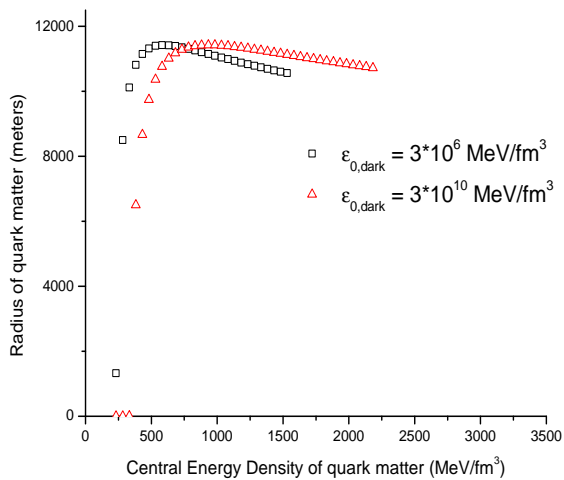}
\caption{Plot of radius (\textit{$R_{quark}$}) vs. central energy density ($\epsilon_{0,quark}$) of quark matter for different central energy densities of dark matter ($\epsilon_{0,dark}$).}
\end{figure}  

\begin{figure}[htb!]
\centering{}
\includegraphics[height=7.5cm,width=3.5in]{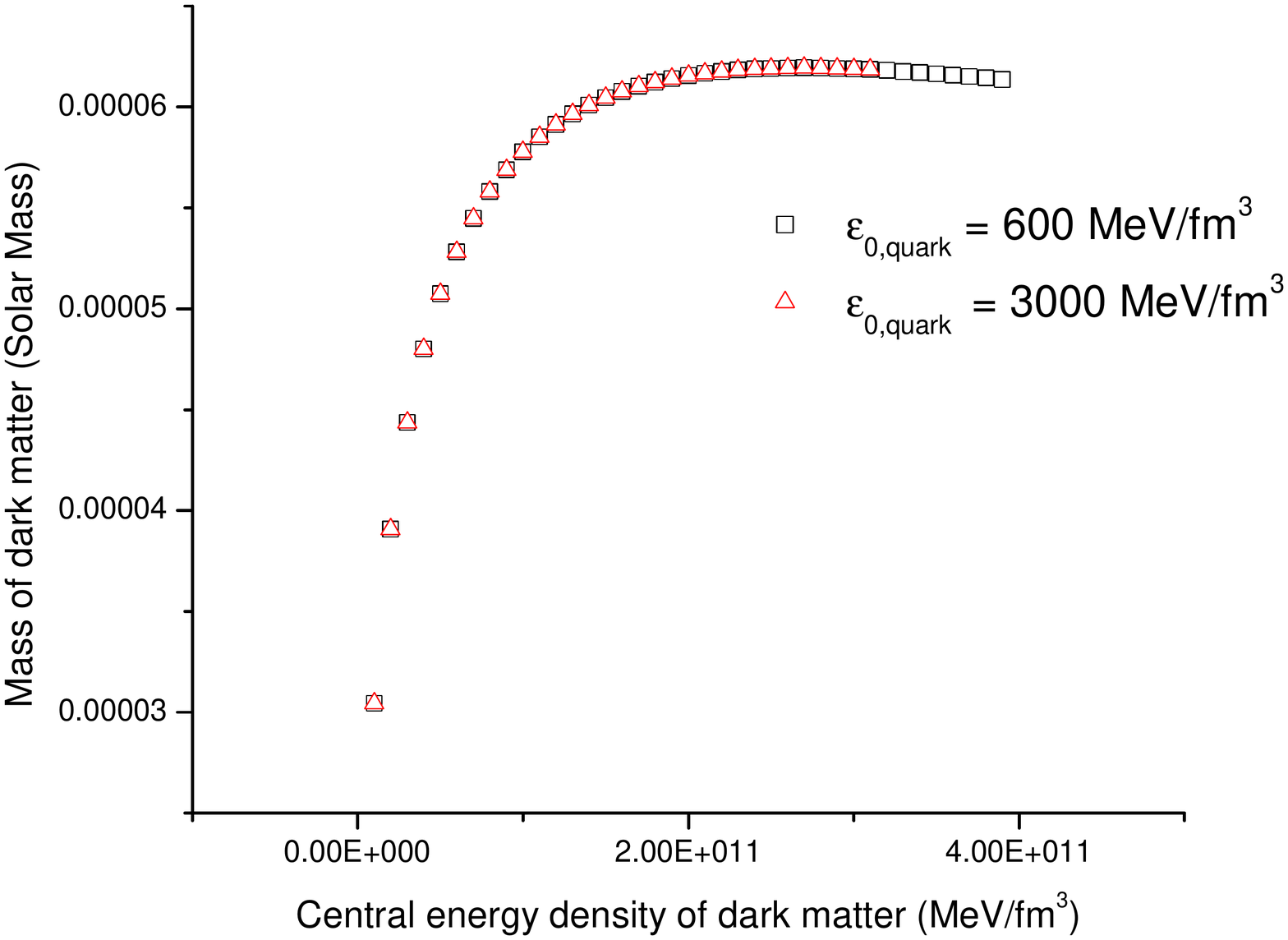}
\caption{Plot of mass (\textit{$M_{dark}$}) vs. central energy density ($\epsilon_{0,dark}$) of dark matter for different central energy densities of quark matter ($\epsilon_{0,quark}$).}
\end{figure} 

\begin{figure}[htb!]
\centering{}
\includegraphics[height=7.5cm,width=3.5in]{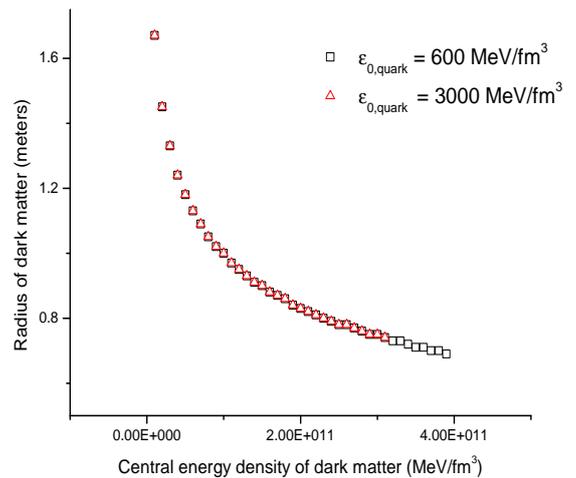}
\caption{Plot of radius (\textit{$R_{dark}$}) vs. central energy density ($\epsilon_{0,dark}$) of dark matter for different central energy densities of quark matter ($\epsilon_{0,quark}$).}
\end{figure}   
Mass (\textit{$M_{quark}$}) vs. central energy density of quark matter ($\epsilon_{0,quark}$) are plotted for three different values of central density of dark matter ($\epsilon_{0,dark}$) each kept constant at a time (See Fig.4). From the plot, it is clear that as the central density of dark matter is increased in the mixture, the maximum mass of the quark matter still reaches to 2.005 $ M_{\odot}$ but now at higher central densities ($\epsilon_{0,quark}$) of quark matter after which the quark matter becomes unstable and the star would collapse. This behaviour can be explained via the fact that as ($\epsilon_{0,dark}$) increases, then within the stable branch of dark matter, the allowed mass of dark matter inside the quark star also increases which contributes to a greater gravitational pull, so, a much higher central quark energy density ($\epsilon_{0,quark}$) is needed to support the maximum possible mass against the greater gravitational pull. The maximum stable mass of the quark component (\textit{$M_{max,quark}$}) is almost the same as pure quark star (2.01 $M_{\odot}$) because the maximum possible value of the dark matter mass is 6.27 . $10^{-5}$ $M_{\odot}$ (Section 5.A) , which is much less than 2.01 $M_{\odot}$, to cause a notable reduction in the quark matter mass. Fig. 5 shows the plot for the radius (\textit{$R_{quark}$}) vs. the central energy density ($\epsilon_{0,quark}$) of quark matter for different values of $\epsilon_{0,dark}$. The figure shows that the maximum stable radius of the quark matter is independent of the amount of free fermionic dark matter present in the admixed star.

\begin{figure}
\centering{}
\includegraphics[height=7.5cm,width=3.5in]{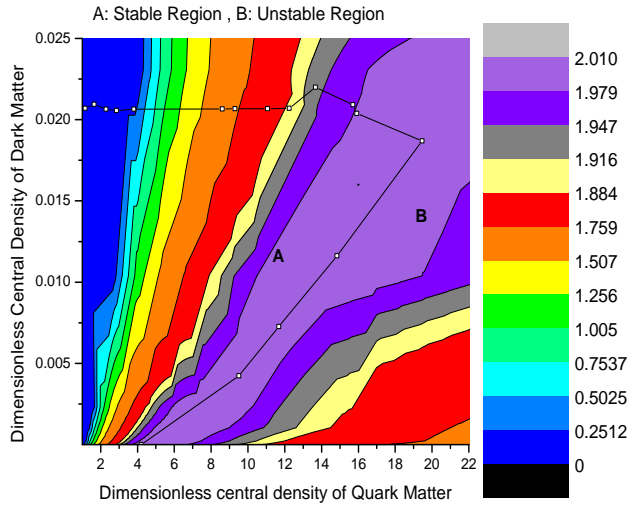}
\caption{Contour plot with (\emph{\(\frac{\epsilon_{0,quark}}{4\textit{B}}\)} , \emph{\(\frac{\epsilon_{0,dark}}{m_f^4}\)} , $M_{total}$(in $M_{\odot}$)  ) as the \textit{x}, \textit{y} and \textit{z} axis respectively. The region marked as A represents the stable region for the formation of quark star admixed with dark matter while all the points in region B are unstable to such a formation. }
\end{figure}

After observing that the maximum possible mass of quark matter is hardly reduced in the presence of free fermionic dark matter particles of various central densities, it is essential to determine which configurations of the admixed star are stable. The plots for the profile of dark matter component is obtained by keeping $\epsilon_{0,quark}$ fixed  and slowly varying $\epsilon_{0,dark}$ (Fig. 6 and 7). It is seen that the dark matter masses and radii are the same for varying $\epsilon_{0,quark}$ which is expected since dark matter is more compact than quark matter and is not affected very much by the presence of quarks.

Figs. 4, 5, 6 and 7 allow us to analyse the stability of the entire configuration. Since we realise from fig. 6 and 7 that $\epsilon_{0,dark}$ for which dark matter mass hits a maxima is the same for all $\epsilon_{0,quark}$, we at first mark those points where the quark matter becomes unstable i.e hits the maximum mass by doing the plots done in figures 4 and 5 for different $\epsilon_{0,dark}$. As we slowly increase $\epsilon_{0,dark}$ , the dark matter content inside the admixed star keeps on increasing and the radius of the dark matter keeps on decreasing. After a sufficiently large $\epsilon_{0,dark}$ , the dark matter mass content hits a maximum after which the dark matter mass decreases with further increase in $\epsilon_{0,dark}$. This is then the unstable branch for the dark matter . After this critical value of $\epsilon_{0,dark}$, no dark matter configurations are stable and hence quark matter and dark matter can't exist together since the dark matter would collapse into a black hole. Hence we expect that up to a certain value of $\epsilon_{0,dark}$ , the quark matter stable mass increases to reach 2.005 $M_{\odot}$ for sufficiently high $\epsilon_{0,quark}$, and after a critical $\epsilon_{0,dark}$, dark matter itself becomes unstable which leads to instability of the entire admixture of the dark matter and quark matter.\\

Next we study the configurations in the $\epsilon_{0,quark}$ - $\epsilon_{0,dark}$ plane. At first, $\epsilon_{0,dark}$ is kept fixed and $\epsilon_{0,quark}$ is slowly increased. The stable boundary is marked in the contour plot (Fig.8) by the maximum stable quark matter mass for increasing $\epsilon_{0,dark}$ which gives the line inclined at an angle in the contour plot. The sequences continue up to the points where the dark matter mass reaches its maxima. Above this value of $\epsilon_{0,dark}$, all configurations become unstable since dark matter itself becomes unstable. This leads to the boundary line that is almost parallel to the x-axis.\\ 
For a quark star admixed with dark matter made of free gas of fermionic particles, the maximum possible mass of the stable configuration is approximately $M_{total}$ $\sim$ 2.01 $M_{\odot}$ with a dark matter content of around 0.63 $\times$ $10^{-4}$ $M_{\odot}$ which has a small radius of about 0.80 meters while the quark matter extends much further to a radius of around 11 km.

\subsection{Solution for combination of quark matter and strongly self-interacting dark matter}

The two fluid TOV equations (eqns. (1), (2), (3), (4)) are solved now for massive dark matter fermions taken to be strongly self-interacting using the model discussed in section 5.B. The interaction strength \textit{y} is taken to be $10^{3}$. The presence of self-interactions causes the maximum stable mass of a dark star to be increased from about $10^{-4}$ \textit{$M_{\odot}$} for the free fermionic case to about $10^{-2}$ \textit{$M_{\odot}$} for the strongly self-interacting case (Section 5.A and 5.B).

\begin{figure}
\centering{}
\includegraphics[height=7.5cm,width=3.5in]{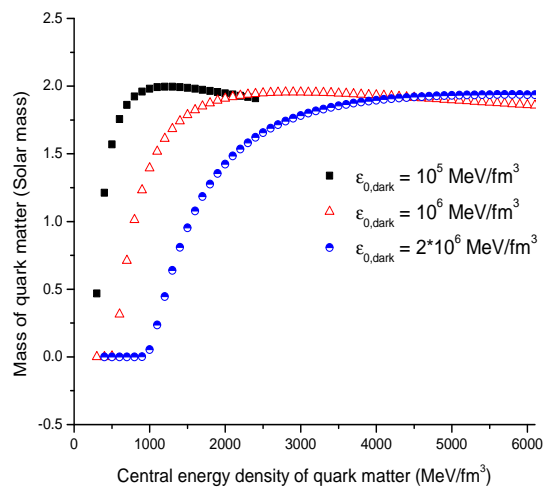}
\caption{Plot for the mass of the quark component ($M_{quark}$) vs. the quark central density $\epsilon_{0,quark}$ for three different dark matter central densities ($\epsilon_{0,intdark}$). It is visible that as $\epsilon_{0,intdark}$ is increased, the maximum stable quark mass ($M_{quark,max}$) is reduced and is now attained at a higher $\epsilon_{0,quark}$ }
\end{figure} 

\begin{figure}
\centering{}
\includegraphics[height=7.5cm,width=3.5in]{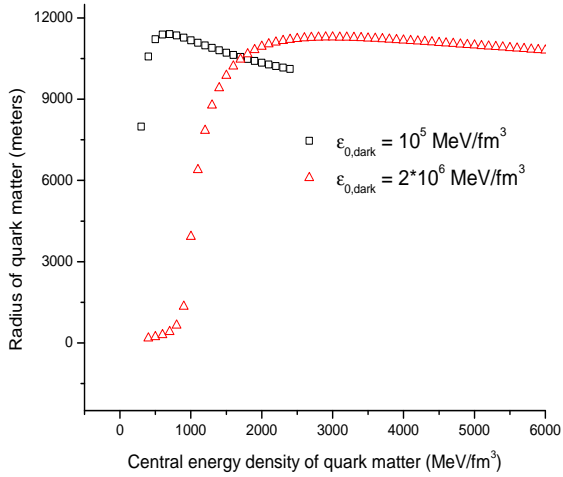}
\caption{Plot for the radius of the quark component ($R_{quark}$) vs. the quark central density $\epsilon_{0,quark}$ for  different dark matter central densities ($\epsilon_{0,intdark}$). }
\end{figure}

The plot for the mass of the quark component (\textit{$M_{quark}$}) vs. the central energy density $\epsilon_{0,quark}$ of the quark component for different values of $\epsilon_{0,intdark}$ (Fig. 9) reveals that the maximum stable mass of the quark matter decreases with increasing central energy density of dark matter ($\epsilon_{0,intdark}$) within the stable branch of dark matter, though the decrease is very moderate. The maximum stable mass of the quark component at $\epsilon_{0,intdark}$ = $10^5$ MeV/$fm^3$ is  1.995 $M_{\odot}$ and this mass reduces to 1.937 $M_{\odot}$ at $\epsilon_{0,intdark}$ = $ 2 \times 10^6 $ MeV/$fm^3$. It is also observed just as in the previous section that the maximum quark mass ($M_{quark,max}$) is attained at a much higher value of $\epsilon_{0,quark}$, the reason being the same as described in free fermion case. The corresponding plot for the radius of the quark component vs. $\epsilon_{0,quark}$ for two different $\epsilon_{0,intdark}$ is shown in Fig. 10 which shows that the maximum radius of quark component also decreases with an increase in $\epsilon_{0,intdark}$ . 
\begin{figure}
\centering{}
\includegraphics[height=7.5cm,width=3.5in]{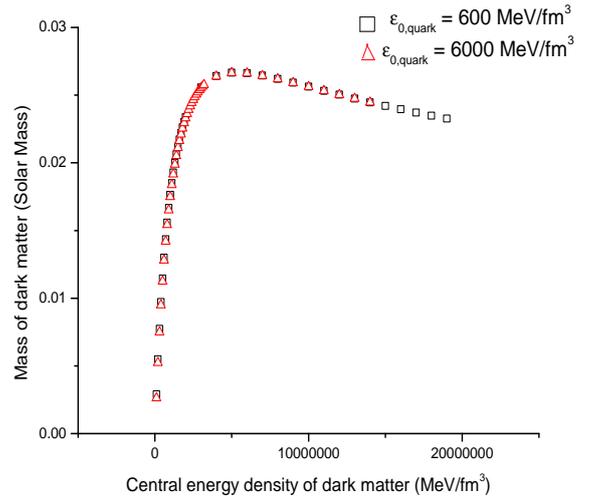}
\caption{Plot for the mass of the dark component ($M_{int}$) vs. the dark central density $\epsilon_{0,intdark}$ for different quark matter central densities ($\epsilon_{0,quark}$). It is visible that the dark matter mass profile is not altered very much with changing $\epsilon_{0,quark}$   }
\end{figure}

\begin{figure}
\centering{}
\includegraphics[height=7.5cm,width=3.5in]{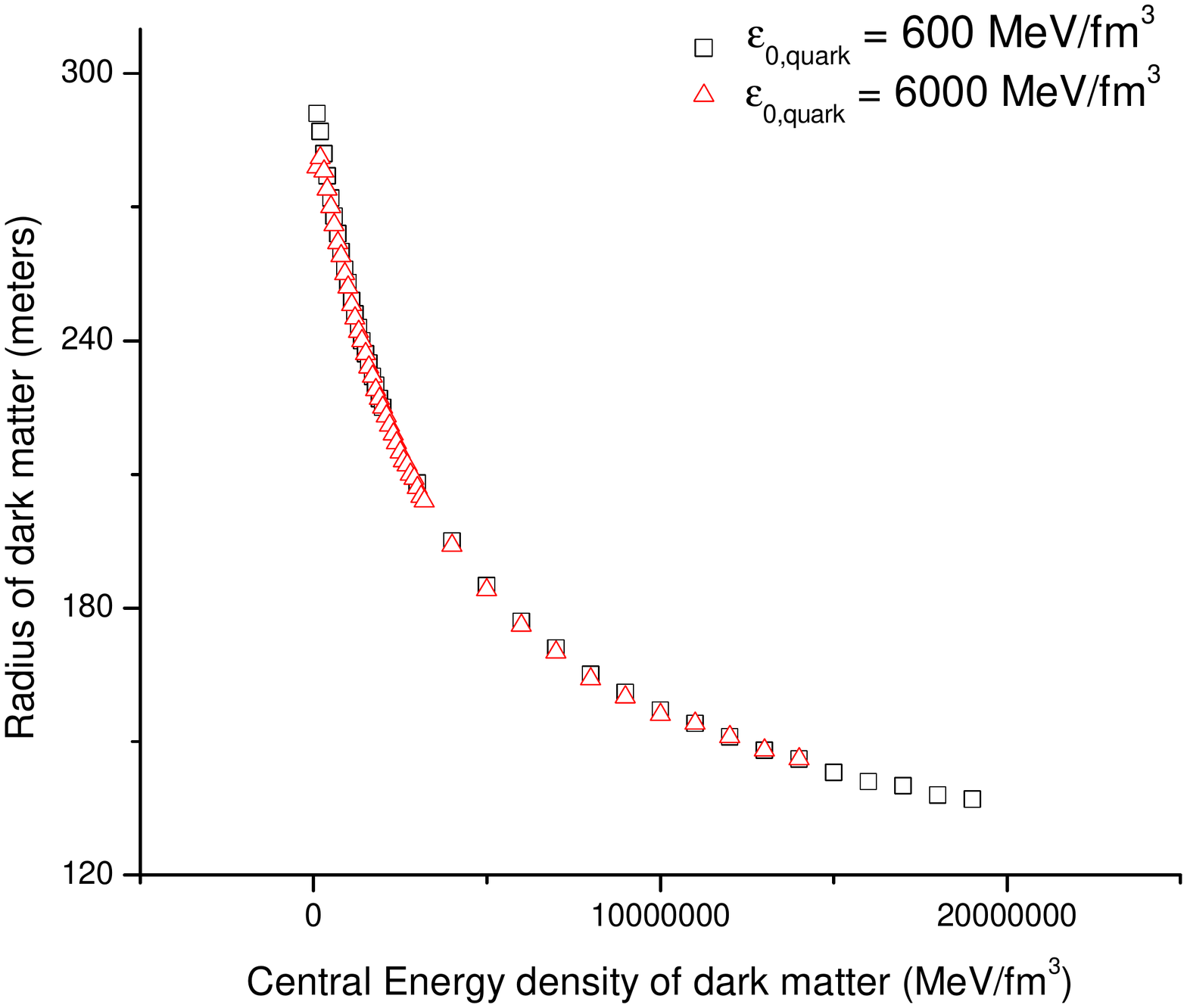}
\caption{Plot for the radius of the dark component ($R_{int}$) vs. the dark central density $\epsilon_{0,intdark}$ for different quark matter central densities ($\epsilon_{0,quark}$). }
\end{figure}

\begin{figure}
\centering{}
\includegraphics[height=7.5cm,width=3.3in]{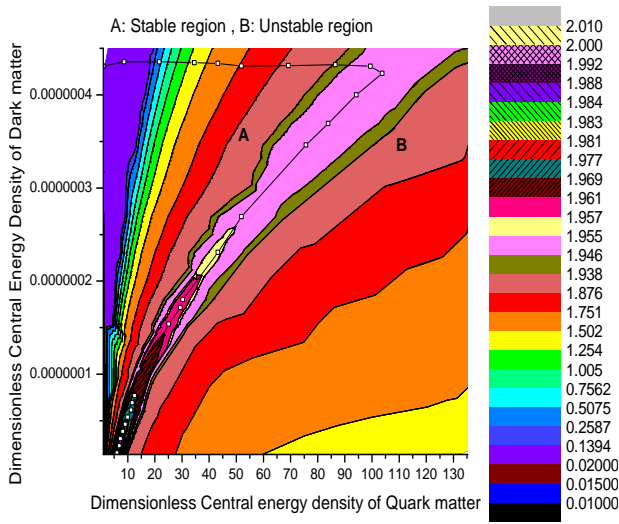}
\caption{Contour plot with (\emph{\(\frac{\epsilon_{0,quark}}{4\textit{B}}\)} , \emph{\(\frac{\epsilon_{0,intdark}}{m_f^4}\)} , $M_{total}$(in $M_{\odot}$)  ) as the \textit{x}, \textit{y} and \textit{z} axis respectively. The region marked as A represents the stable region for the formation of quark star admixed with dark matter while all the points in region B are unstable to such a formation. \emph{\(\frac{\epsilon_{0,quark}}{4\textit{B}}\)} starts from 1.0 in the plot because energy density of quark matter can't be zero according to the equation of state (6). }.
\end{figure}  

Keeping $\epsilon_{0,quark}$ constant and obtaining $M_{int}$ vs. $\epsilon_{0,intdark}$ and $R_{int}$ vs. $\epsilon_{0,intdark}$  gives the profile for dark matter present in the admixture (Fig. 11 and 12). It is evident from Figs. 11 and 12 that the dark matter mass and radius profile does not change much with increasing $\epsilon_{0,quark}$ since dark matter is much more compact than quark matter and its particles are also much more massive to be significantly affected by quark matter particles. Figs. 9, 10, 11 and 12 allow us to determine the stability of the quark matter star admixed with self-interacting dark matter. For the first two plots showing the dependence of the mass and radius of the quark matter vs. $\epsilon_{0,quark}$ tells us up to which point the quark star configuration would remain stable by noting the point of maxima of the mass and the radius. The next two plots , Fig. 11 and 12 allows us to determine up to which point the dark matter remains stable for varying $\epsilon_{0,quark}$ . It is evident from the graphs that the dark matter parameter profile is almost independent of $\epsilon_{0,quark}$. So, the $\epsilon_{0,intdark}$  at which the dark matter becomes unstable is the same for all $\epsilon_{0,quark}$. 

The contour plot showing the dependence of the total mass of the entire star ($M_{total}$) on the dimensionless central energy densities of the two fluids (Fig. 13) reflects the decrease in the maximum stable total mass with increase in $\epsilon_{0,intdark}$.

The region of stability is marked in the contour diagram . The shape of the boundary is similar to the free case discussed before. The upper branch of the boundary line is an indicator of the $\epsilon_{0,intdark}$ after which the the dark matter component becomes unstable for a given $\epsilon_{0,quark}$. As a quick check, in the contour diagram, the plot converges to the appropriate mass limit for low  $\epsilon_{0,intdark}$. For low $\epsilon_{0,intdark}$ say, $10^5$ MeV/$fm^3$, the maximum stable mass is $\sim$ 2.0 $M_{\odot}$ at a radius of about 11 km showing the convergence to pure quark star limit.

\begin{figure}
\centering{}
\includegraphics[height=7.5cm,width=3.5in]{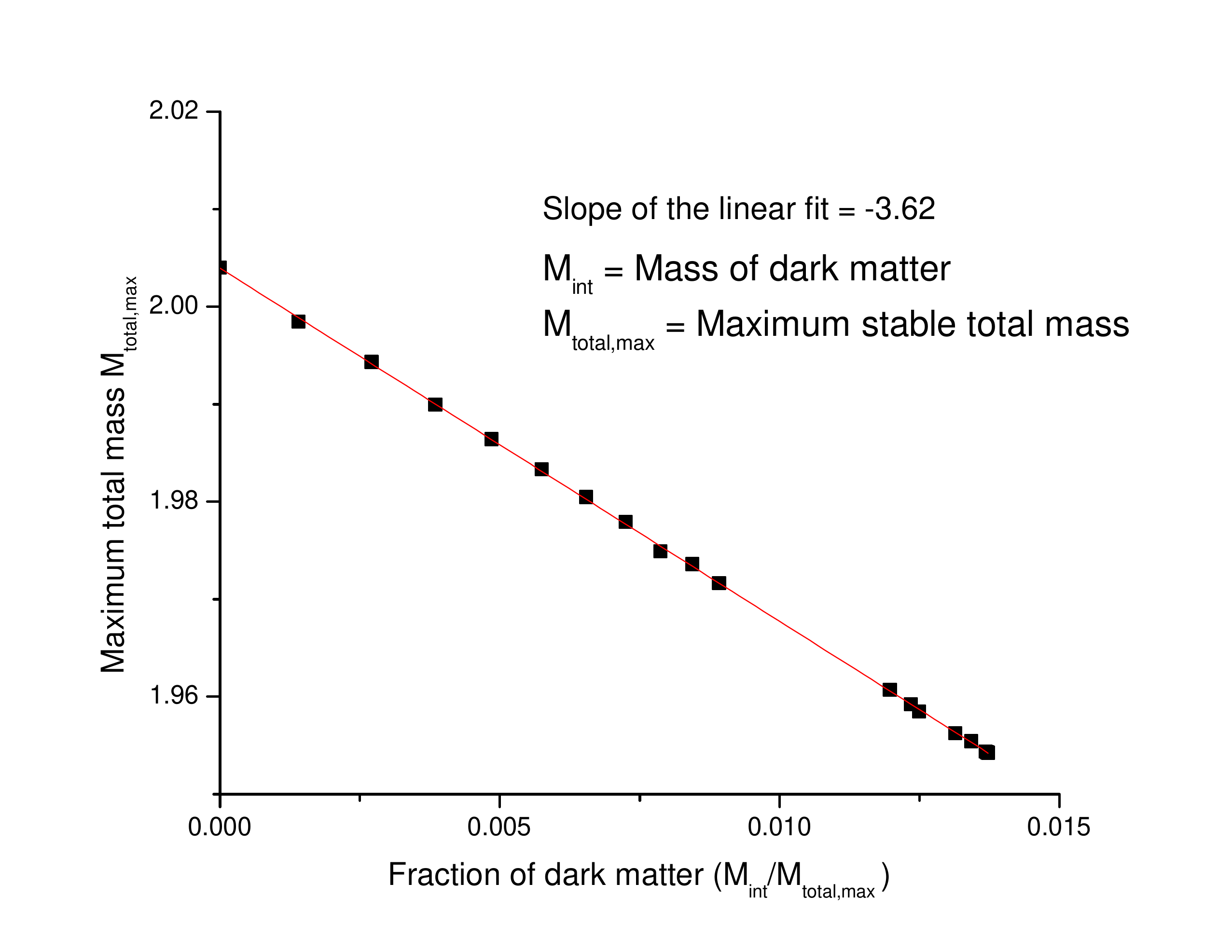}
\caption{ Plot showing the dependence of maximum total mass vs. the dark matter content for quark star admixed with dark matter. The slope of the plot is -3.62. It shows that the maximum total mass decreases with increasing dark matter content in the star. The red line shows the linear fit.}
\end{figure}

Fig. 14 shows the maximum total mass vs. the fraction of dark matter which can be fitted by a linear fit. The slope of the fit comes out to be about -3.62.

The reason for the decrease of the maximum total mass with increasing dark matter content is the increased gravitational force due to extra dark matter content which causes a collapse of the star. Using the linear fit in the Fig. 14, the equation for the dependence of the maximum total stable mass of the admixed star on the dark matter fraction present is given as:

\begin{equation}
\small
\frac{M_{tot,max}}{M_{\odot}} = 2.004- 3.62 \emph{\(\frac{M_{int}}{M_{tot,max}}\)}
\end{equation}
or,
 
\begin{equation}
\small
\frac{M_{tot,max}}{M_{\odot}} = 2.004- 3.62 f
\end{equation}

where \textit{f} is the fraction of self-interacting dark matter in the admixed star at the maximum stable total mass. The maximum allowed strongly self-interacting dark matter content is about 2.64 $\times$ $10^{-2}$ $M_{\odot}$ at a maximum total stable mass of about 1.95 $M_{\odot}$ which gives a maximum limit on the possible dark matter fraction $f_{max}$ $\simeq$ 0.014. Radio timing observations of the pulsar J0348+0432 and phase-resolved optical spectroscopy of its white-dwarf companion lead to a precise pulsar mass measurement of 2.01 $\pm$ 0.04 $M_{\odot}$ which is by far the highest yet measured with this precision \cite{Demorest:2010abc,Antoniadis:2013abc}. The maximum stable dark admixed quark star mass (1.95 $M_{\odot}$) falls slightly lower than the error limit of the highest measured pulsar mass.

\section{Summary and Discussions}

Pure quark matter is studied by using the MIT Bag model \cite{Alcock:1986abc}. Dark matter stars are studied here by considering the dark matter to be made up of fermionic particles of mass 100 GeV \cite{Kaplan:1992abc,Nussinov:1985abc} with the assumption that these particles do not self-annihilate. The maximum stable mass of the dark matter star composed of strongly self-interacting particles is about 2.7$\times$ $10^{-2}$ $M_{\odot}$ at a radius of about 0.19 km while for free fermions, the mass is about 6.0$\times$ $10^{-5}$ $M_{\odot}$ at a radius of around 1 meter. \\ 
The complete dimensional two-fluid TOV equations equations are solved to study the behaviour of admixed quark matter with dark matter. First, the equations are solved for a mixture of quark matter and free dark matter. The maximum stable mass of the admixed star is almost the same as that for a pure quark star for increasing dark matter fraction within the star. As the content of dark matter is gradually increased in the admixed star, the dark matter reaches its maximum stable configuration after which no admixed star configuration remains stable since the dark matter component collapses to form a black hole. For a quark star admixed with dark matter made of free gas of particles, the maximum possible mass of the stable configuration is approximately $M_{total}$ $\sim$ 2.01 $M_{\odot}$ with a dark matter content of around 0.63 $\times$ $10^{-4}$ $M_{\odot}$. A reduction in the maximum stable total mass is noted in case of a quark star admixed with dark matter star composed of strongly self-interacting fermions. The decrease was from about 2.01 $ M_{\odot}$ for zero dark matter content inside the star to about 1.95 $ M_{\odot}$ for the maximum allowed mass of strongly interacting dark matter in the star. The maximum dark matter content is around 2.64 $\times$ $10^{-2}$ $M_{\odot}$ at a maximum stable total mass of about 1.95 $M_{\odot}$. The maximum stable total mass in case of strongly self-interacting dark matter is seen to reduce linearly with increasing dark matter fraction in the star. The maximum accretion rate of dark matter by the quark star can be estimated to be about \emph{\(\frac{M_{int,max}}{\tau}\)} $\sim$ 2.03$\times$ $10^{-12}$ $M_{\odot}$ per year, where $\tau$ $\sim$ 1.3 $\times$ $10^{10}$ years is the estimate for the age of the universe and $M_{int,max}$ is the maximum possible self-interacting dark matter content in the quark star. If the accretion rate is higher than this, the quark star will collapse .  \\
\section{Acknowledgements}     
This work started started as a summer project of P.M at the Goethe University in Frankfurt and is supported by Deutscher Akademischer Austausch Dienst (DAAD). P.M thanks the Institute of Theoretical Physics for their hospitality. We thank Andreas Zacchi, Rainer Stiele and Chhanda Samanta for helpful discussions and a critical reading of the manuscript.\\


\begin{thebibliography}{40}
\bibliographystyle{abbrv}


\bibitem{Witten:1984abc}
  E.~Witten,
  Phys.\ Rev.\ D {\bf 30} (1984) ,  272.

\bibitem{Itoh:1970abc}
  N.~Itoh, Prog.
   Theor.\ Phys. {\bf 44} (1970) 291.

\bibitem{Farhi:1984abc}
  Farhi~E. and Jaffe~R.~L.,
  Phys.\ Rev.\ D {\bf 30} (1984).

\bibitem{Weber:2005abc}
  F.~Weber, Prog.~Part.
  Nucl.\ Phys. {\bf 54} (2005) 193.

\bibitem{Ivanenko:1965abc}
  D.~D.~Ivanenko, D.~F.~Kurdgelaidze,
  Astrophysics {\bf 1} (1965).

\bibitem{Chodos:1974abc}
  Chodos~A., Jaffe~R.~L., Johnson~K., Thorne~C.~B. and Weisskopf~V.~F.,
  Phys.\ Rev.\ D {\bf 9} (1974).
  
  
\bibitem{Haensel:1986abc}
  P.~Haensel, J.~L.~Zdunik and R.~Schaeffer,
  Astron.\ Astrophys. {\bf 160} (1986) 121.

\bibitem{Alcock:1986abc}
  C.~Alcock, E.~Farhi and A.~Olinto,
  Astrophys.\ J. {\bf 310} (1986) 261.

\bibitem{Baltz:2004abc}
  E.~A.~Baltz, 
  Proceedings of the 32nd SLAC Summer Institute on Particle Physics (SSI 2004): Natures Greatest Puzzles, Menlo Park, California, 2004, eConf {\bf C040802}, L002 (2004).

\bibitem{Kaplan:1992abc}
  D.~Kaplan, 
  Phys.\ Rev.\ Lett {\bf 68} (1992) 741.

\bibitem{Nussinov:1985abc}
  S.~Nussinov, 
  Phys.\ Lett.\ B {\bf 165} (1985) 55.

\bibitem{Perez:2010abc}
  M.~A.~Perez~-Garcia, J.~Silk. and J.~R.~Stone,
  Phys.\ Rev.\ Lett {\bf 105} (2010) 141101.

\bibitem{Perez:2012abc}
  M.~A.~Perez~-Garcia and J.~Silk, 
  Phys.\ Lett.\ B {\bf 711} (2012) 6.

\bibitem{Kouvaris:2011abc}
  C.~Kouvaris and P.~Tinyakov,
  Phys.\ Rev.\ D {\bf 83} (2011) 083512.

\bibitem{Goldman:2013qla}
  I.~Goldman, R.~N.~Mohapatra, S.~Nussinov, D.~Rosenbaum and V.~Teplitz,
  Phys.\ Lett.\ B {\bf 725} (2013) 200
  [arXiv:1305.6908 [astro-ph.CO]].

\bibitem{Kouvaris:2015rea}
  C.~Kouvaris and N.~G.~Nielsen,
  Phys.\ Rev.\ D {\bf 92} (2015) 6,  063526
  [arXiv:1507.00959 [hep-ph]].

\bibitem{Lavallaz:2010abc}
  A.~de~Lavallaz and M.~Fairbairn,
  Phys.\ Rev.\ D {\bf 81} (2010) ,  123521.
  
\bibitem{Li:2012abc}
  A.~Li and F.~Huang and R.~X.~Xu,
  Astropart.\ Phys. {\bf 37} (2012) 70.

\bibitem{Leung:2011abc}
  S.~C.~Leung and M.~C.~Chu and L.~M.~Lin,
  Phys.\ Rev.\ D {\bf 84} (2011) ,  107301.

\bibitem{Sandin:2009abc}
  F.~Sandin and P.~Ciarcelluti,
  Astropart.\ Phys. {\bf 32} (2009) 278.
  
\bibitem{Xiang:2014abc}
  Q.~F.~Xiang, W.~Z.~Jiang, D.~R.~Zhang and R.~Y.~Yang,
  Phys.\ Rev.\ C {\bf 89} (2014) , 025803.

\bibitem{Laura:2015abc}
  Laura ~Tolos and J.~Schaffner.~Bielich,
  [arXiv:1507.08197 [astro-ph.HE]] (2015).
  
\bibitem{Fraga:2001abc}
  E.~S.~Fraga, R.~D.~Pisarski and J.~Schaffner.~Bielich, 
  Phys.\ Rev.\ D {\bf 63} (2001).
  
\bibitem{Sagert:2006abc}
  I.~Sagert, M.~Hempel and C.~Greiner and J.~Schaffner-~Bielich ,
  Eur.\ J.\ Phys. {\bf 27} (2006) 577.

\bibitem{Macher:2005abc}
  J.~Macher and J.~Schaffner.~Bielich,
  Eur.\ J.\ Phys {\bf 26} (2005) 341.

\bibitem{Silbar:2004abc}
  R.~R.~Silbar and S.~Reddy,
  Am.\ J.\ Phys {\bf 72} (2004) 892; {\bf 73} (2005) 286(E).

\bibitem{Haensel:2007abc}
  P.~Haensel, A.~Y.~Potekhin and D.~G.~Yakovlev,
  New\ York\ : Springer, page 2 (2007).

\bibitem{Gaurav:2006abc}
  Gaurav~Narain, J.~Schaffner.~Bielich and Igor~N.~Mishustin,
  Phys.\ Rev.\ D {\bf 74} (2006).
  
\bibitem{Pratik:2009abc}
  Pratik~Agnihotri, J.~Schaffner.~Bielich and Igor~ N.~Mishustin,
  Phys.\ Rev.\ D {\bf 79} (2009).

\bibitem{Rainer:2010abc}
  Rainer~Stiele, Tillman~Boeckel and J.~Schaffner.~Bielich, 
  Phys.\ Rev.\ D {\bf 81} (2010).

\bibitem{Bottino:2005abc}
  A.~Bottino, F.~Donato, N.~Fornengo and S.~Scopel,
  Phys.\ Rev.\ D {\bf 72} (2005).
  
\bibitem{Andreas:2015abc}
  Andreas~Zacchi, Rainer~Stiele and J.~Schaffner.~Bielich, 
  Phys.\ Rev.\ D {\bf 92} (2015).

\bibitem{Demorest:2010abc}
  P.~Demorest, T.~Pennucci, S.~Ransom, M.~Roberts and J.~Hessels,
  Nature {\bf 467} (2010) 1081.

\bibitem{Antoniadis:2013abc}
  J.~Antoniadis, P.~C.~C.~Freire, N.~Wex, T.~M.~Tauris, R.~S.~Lynch, M.~H.~van~Kerkwijk, M.~Kramer and C.~Bassa, 
  Science {\bf 340} (2013) 6131.




 
\end{thebibliography}
\end{document}